\documentclass[letterpaper,11pt]{article}
\usepackage[hmargin=1in,vmargin=1in]{geometry}
\usepackage{amsmath}
\usepackage{graphicx}
\usepackage{bm}
\usepackage{epsfig}
\title{On the gravitational instability in the Newtonian limit of MOG}
\author{Fatimah Shojai$^{1,2}$, Samira Cheraghchi$^1$,  Hamed Bouzari Nezhad$^1$ \\ $^1$Department of Physics, University of Tehran,\\ Tehran, Iran.\\$^2$Foundations of Physics Group, School of Physics,\\ Institute for Research in Fundamental Sciences (IPM),\\ Tehran, Iran.\\}
\date{}
\begin{document}
\maketitle
\begin{abstract}
We have found some analytical cosmological solutions to MOdified Gravity (MOG). These solutions describe different evolutionary epochs of an isotropic and homogeneous universe. During each epoch, the evolution of cosmological perturbation is studied in the Newtonian framework and compared with the corresponding results of GR. 
\end{abstract}
\section{Introduction}
There are many observations in cosmology leading to the introduction of mysterious aspects of dark matter and dark energy. One can think of an alternative gravitational dynamics instead of introducing dark matter and dark energy. At present, many modified gravitational theories are proposed \cite {koli}. Two usual approaches are adding higher order curvature invariants and extra  fields in the gravitational action. For the latter case, two important extended theories are Tensor-Vector-Scalar (TeVeS) theory \cite{tev} and MOdified Gravity (MOG) \cite{moffat}. In both, gravity is described by some tensor, vector and scalar fields and there exist some free parameters which must be fitted to observations. 

MOG introduced in order to explain the flat rotation curves of spiral galaxies and mass discrepancy in galaxy clusters without the need of exotic dark matter \cite{rotationcurves} as well as explaining the large scale structure of the universe \cite{moffatarxiv2014}. In this article we study how small initial inhomogeneities  grow in an expanding universe in MOG. We take the Newtonian viewpoint which is an adequate description of relativistic treatment on sub-horizon scale and for non-relativistic matter perturbation.
To do this, it is required to have the background metric of space--time. Thus we must first find some cosmological solutions of MOG.  Some of these solutions are obtained in \cite {mof} using the numerical methods and in \cite{roshan} via the Noether symmetry approach \cite{nsa}. 

In this paper, after reviewing the basic equations of MOG, we derive some cosmological solutions for a spatially flat Friedmann--Robertson--Walker (FRW) universe with a perfect fluid in section 3. The first one corresponds to an exact power-law evolution of dynamical fields while the other one corresponds to a universe which is dominated by a single component fluid together with $G$-field.  We consider $G$-radiation, $G$-phion and $G$-matter dominated universes. These are interested since we want to study the evolution of inhomogeneity in these epochs. A brief discussion of Newtonian analysis of gravitational instability in MOG \cite{nagimog} is presented in section 4. Then we study how the sub-Hubble fluctuations evolve in an expanding universe and compare the result with standard cosmology in section 5.

\section{The cosmological field equations of MOG}
MOG theory postulates more gravitational fields than GR. In addition to metric tensor, $g_{\alpha\beta}$, there are a massive Proca vector field $\phi_\mu(x)$ which is coupled to matter  and two additional scalar fields, $\mu(x)$, the mass of the vector field,  and $G(x)$, a variable  gravitational constant. The action of this theory can be written as \cite{moffat}:
\begin{align}
S=S_{grav}+S_\Lambda+S_\phi+S_\mu+ S_G+ S_M
\end{align}
where  $S_M$, is the matter sector of action and  $S_{grav}$, $S_\Lambda$,  $S_\phi$,  $S_\mu$ and  $S_G$  are given by:
\begin{align}
S_{grav}=\frac{1}{16\pi}\int \sqrt{-g}\frac{1}{G} R d^4x
\end{align}
\begin{align}
S_\Lambda=-\frac{1}{8\pi}\int \sqrt{-g}\frac{1}{G}\Lambda d^4x
\end{align}
\begin{align}
S_\phi=
-\int\sqrt{-g}\omega_0\left\lbrack\frac{1}{4}B_ {\alpha\beta}B^{\alpha\beta}
+V_\phi\right\rbrack~d^4x
\end{align}
\begin{align}
S_\mu=
\int\sqrt{-g}\frac{1}{G\mu^2}\left\lbrack\frac{1}{2}g^{\alpha\beta}\nabla_\alpha\mu\nabla_\beta\mu
-V_\mu\right\rbrack~d^4x
\end{align}
\begin{align}
S_G=
\int\sqrt{-g}\frac{1}{G^3}\left\lbrack\frac{1}{2}g^{\alpha\beta}\nabla_\alpha G\nabla_\beta G
-V_G\right\rbrack~d^4x
\end{align}
in which $\omega_0$ is a dimensionless positive coupling constant and $B_{\alpha\beta}=\nabla_\alpha\phi_\beta-\nabla_\beta\phi_\alpha$. $R$ is the Ricci scalar and $\Lambda$ is the cosmological constant. $V_\phi$, $V_\mu$ and $V_G$ are self interaction potentials of the vector and scalar fields. The constant coupling parameter $\omega_0$ in the action and the resulting field equations plays no significant role and can be absorbed in other variables, and thus hereafter we set it equal to unity. Considering a spatially flat FRW  metric:
\begin{align}
dS^2=-dt^2+a^2(t)\left (dr^2+r^2d\Omega^2 \right )
\end{align}
in which $a(t)$ is the scale factor of the universe. Here we shall use the original MOG potential in the form \cite{moffat}:
\begin{align}\label{mogpotential}
V_\phi=-\frac{1}{2}\mu^2\phi_\mu\phi^\mu
\end{align}
 and set the other potential functions to zero.
The MOG cosmological equations are:
\begin{align}\label{eqn1}
\frac{\dot{a}^2}{a^2}~&=~\frac{8\pi G\rho}{3}+\frac{\Lambda}{3}+\frac{\dot{G}}{G}\frac{\dot{a}}{a}-\frac{1}{12}\frac{\dot{\mu}^2}{\mu^2}-\frac{1}{24}\frac{\dot{G^2}}{G^2}+\frac{G}{12}\mu^2\phi_0^2  
\end{align}
\begin{align}\label{eqn2}
\frac{\ddot{a}}{a}~&=~\frac{-4\pi G}{3}(\rho+3p)+\frac{\Lambda}{3}+\frac{1}{2}\frac{\dot{G}}{G}\frac{\dot{a}}{a}+\frac{1}{6}\frac{\dot{\mu}^2}{\mu^2}+\frac{1}{2}\frac{\ddot{G}}{G}-\frac{11}{12}\frac{\dot{G}^2}{G^2}-\frac{G}{6}\mu^2\phi_0^2   
\end{align}
\begin{align}\label{eqn3}
\frac{\ddot{G}}{G}=32\pi G\rho+12\frac{\ddot{a}}{a}+9\frac{\dot{G}}{G}\frac{\dot{a}}{a}-2\frac{\dot{\mu}^2}{\mu^2}+\frac{\dot{G}^2}{G^2}+G\mu^2\phi_0^2    
\end{align}
\begin{align}\label{eqn4}
\frac{\ddot{\mu}}{\mu}=\frac{\dot{\mu}^2}{\mu ^2}-3\frac{\dot{\mu}}{\mu}\frac{\dot{a}}{a}+\frac{\dot{G}}{G}\frac{\dot{\mu}}{\mu}-G\mu\frac{\partial V_{\phi}}{\partial \mu}
\end{align}
\begin{align}\label{eqn5}
\frac{\partial V_{\phi}}{\partial\phi_0}=16\pi\kappa\rho
\end{align}
where $\kappa$ is a coupling constant that appears in variation of matter action with respect to $\phi_\alpha$ \cite{moffat} and a dot denotes derivative with respect to the cosmic time. The zeroth component of the vector field is the only non-zero component of $\phi$-field because of the cosmological principle which also implies the conservation of energy-momentum tensor of cosmological fluid in MOG theory \cite{roshantp}. The above equations are generalized Friedmann equations and the equations of motion for $G$ and $\mu$ fields respectively. The last relation gives the coupling of matter to $\phi$ field which is neutral and doesn't couple to photons. Therefore, $\phi$ field perturbations can grow during the radiation dominated era in which baryons and photons are strongly coupled.

Using the definition of the energy-momentum tensor of fields (the index $f$ can be $\Lambda$, $\phi$, $\mu$ or $G$.) 
\begin{align}
T_{\mu\nu}=\frac{-2}{\sqrt{-g}}\frac{\delta S_f}{\delta g^{\mu\nu}}
\end{align}
we have:
\begin{eqnarray}\label{phitensor}
T^{(\phi)}_{00}=\rho^{(\phi)}=\frac{1}{32\pi}\mu^2\phi_0^2
\end{eqnarray}
\begin{eqnarray}\label{mutensor}
T^{(\mu)}_{00}=\rho^{(\mu)}=-\frac{1}{32\pi G}\frac{\dot\mu^2}{\mu^2}
\end{eqnarray}
\begin{eqnarray}\label{gtensor}
T^{(G)}_{00}=\rho^{(G)}=-\frac{1}{64\pi G}\frac{\dot G^2}{G^2}
\end{eqnarray}
\begin{eqnarray}\label{lambdatensor}
\rho^{(\Lambda)}=\frac{\Lambda}{8\pi G}
\end{eqnarray}
Using the above relations, one can rewrite the first two equations of motion, (\ref{eqn1}) and (\ref{eqn2}), as follows:
\begin{eqnarray}\label{eeqn1}
\frac{\dot{a}^2}{a^2}=\frac{8\pi G}{3}\big(\rho+\rho^{(\Lambda)}+\rho^{(\phi)}+\rho^{(\mu)}+\rho^{(G)}\big)+\frac{\dot{G}}{G}\frac{\dot{a}}{a}
\end{eqnarray}
\begin{eqnarray}\label{eeqn2}
\begin{split}
\frac{\ddot{a}}{a}=&-\frac{4\pi G}{3}\Big[(\rho+3 p)+\rho^{(\Lambda)}(1+3\omega_\Lambda)+\rho^{(\phi)}(1+3\omega_\phi)+\rho^{(\mu)}(1+3\omega_\mu)\\&+\rho^{(G)}(1+3\omega_G)\Big]+\frac{1}{2}\frac{\ddot G}{G}+\frac{1}{2}\frac{\dot G}{G}\frac{\dot a}{a}-\frac{\dot G^2}{G^2}-\frac{1}{8}G\mu^2\phi_0^2  
\end{split}
\end{eqnarray}
where $\omega_\Lambda=-1$, $\omega_\mu=\omega_G=1$, $\omega_\phi=0$.
We use this form of equations in the next section in which we study how a single or two fluid components drive the evolution of the universe.

To obtain a closed system of equations, one must specify one equation of state for matter which is usually assume to be linear in cosmology:
\begin{align}
 p=\omega\rho
\end{align}
where the equation of state parameter $\omega$ is a constant.

Before discussing some cosmological solutions of MOG theory, let us explain briefly how this theory is consistent with the observational data. Below, we list some of the more important ones:
\begin{itemize}

\item{Big-Bang nucleosynthesis

At the nucleasynthesis era, the gravitational constant is very close to the Newtonian one \cite{moffatarxiv2014} and thus the abundances of the light elements agree with the available data.}

\item{Rotation curve

Using the weak field approximation of MOG \cite{rah1,rah2}, one obtains an effective gravitational potential which has two free parameters, $\alpha=(G_\infty-G_N)/G_N$ and $\tilde{\mu}$. $G_N$ is the Newtonian gravitational constant, $G_\infty$ is the effective gravitational constant at infinity and $\tilde{\mu}$ is the mass of vector field which is constant at this approximation. Best fitting of the galaxy rotatoin curves gives:
$\alpha\simeq8.89\pm0.34$ , $\tilde{\mu}\simeq0.042\pm0.004$ ${kpc}^{-1}$ \cite{rah1,rah2}.}
\item{Baryon acoustic oscillations

Since the Jeans length of pressureless phion particles (the particle of $\phi$ field) is very small, there is no oscillatory behavior for these particles. Their perturbations grow while baryon perturbations oscillate before decoupling \cite{moffatarxiv2014}.}

\item{The CMB power spectrum

In the present universe:
$(\Omega_b^0)_{MOG}=(\Omega_b^0)_{\Lambda CDM}$, $\rho_{\phi}^0\ll\rho_{b}^0$, $G^0=G_N(1+\alpha)$ and $(G_N\rho)_{\Lambda CDM} =(G_N(1+\alpha)\rho)_{MOG}$ where $\rho_{\Lambda CDM}=\rho_b+\rho_{CDM}$ and $\rho_{MOG}=\rho_b$. Going  back to the past, at the decoupling time, $\rho_{\phi}\gg\rho_b$ and $\alpha\ll1$, thus $(G_N\rho)_{\Lambda CDM}=(G_N\rho_{\phi})_{MOG}$ \cite{moffatarxiv2014,moffatarxiv2015}. This shows that the CMB power spectrum in MOG agrees with the corresponding one in $\Lambda CDM$ model.}

\end{itemize}

\section{Some exact cosmological solution of MOG}

In this section we find some cosmological solutions of MOG theory with zero cosmological constant. The simplest analytical solutions are power-law type. Assuming that:
\begin{align}
\label{q}
a(t)=(\frac{t}{t_0})^\lambda,\hspace{0.2in}G(t)=G_0(\frac{t}{t_0})^\sigma,\hspace{0.2in}\rho(t)=\rho_0(\frac{t}{t_0})^{-3(1+\omega)\lambda},\hspace{0.2in}\mu(t)=\mu_0(\frac{t}{t_0})^\alpha,\hspace{0.2in}\phi_0(t)=\tilde{\phi_0}(\frac{t}{t_0})^\beta
\end{align}
where a subscript "0" indicates the value of any quantity evaluated at the present and the usual normalization $a_0=1$ is used. Inserting (\ref{q}) into equations (7)-(12), we find that
\begin{align}
\sigma=\frac{2(-5+30\omega+3\omega^2)}{23-6\omega+3\omega^2}
\end{align}
\begin{align}
\lambda=\frac{4(3+\omega)}{23-6\omega+3\omega^2}
\end{align}
 \begin{align}
\alpha=-\frac{6(3+4\omega+\omega^2)}{23-6\omega+3\omega^2}
\end{align}
\begin{align}
\rho_0=-\frac{9\mu_0^2(3+4\omega+\omega^2)}{16\pi\kappa^2(-34+9\omega+3\omega^2)}
\end{align}
\begin{align}
\tilde{\phi_0}=-\frac{3(3+4\omega+\omega^2)}{\kappa(-34+9\omega+3\omega^2)}
\end{align}
\begin{align}
G_0=-\frac{2\kappa^2(-34+9\omega+3\omega^2)^2(-23+42\omega+9\omega^2)}{3\mu_0^2(3+4\omega+\omega^2)(23-6\omega+3\omega^2)^2}
\end{align}
\begin{align}
\beta=0
\end{align}
For dust, $\omega=0$ and the exact solution is simplified to the following   form:
\begin{align*}\label{w0}
a(t)=(\frac{t}{t_0})^{12/23}
\end{align*}
\begin{align*}
\rho_m(t)=\frac{9}{544}\frac{{\mu_0}^2}{\pi\kappa^2}(\frac{t}{t_0})^{-36/23}
\end{align*}
\begin{align}
G(t)=\frac{2312\kappa^2}{207{\mu_0}^2}(\frac{t}{t_0})^{-10/23}
\end{align}
\begin{align*}
\mu(t)=\mu_0(\frac{t}{t_0})^{-18/23}
\end{align*}
\begin{align*}
\phi_0(t)=\frac{9}{34\kappa}
\end{align*}
This solution is recently obtained in \cite{roshan} using the Noether symmetry approach. Note that a property of this solution is that all types of energy densities, (\ref{phitensor})-(\ref{gtensor}), evolve in a similar way, proportional to $a(t)^{-3}$. This solution corresponds to the epoch in which two pressure-less components, phion and baryon, present and participate in structure formation.

Now looking at other simplified solutions, in which there is a cosmological fluid and $G$-field density term on the right hand side of equation (\ref{eeqn1}). The reason why we consider these, requires some words of explanation.
The cosmological evidences indicate that we live in a radiation dominated universe during the early stages. In that era, photons and baryons form a tightly coupled fluid and the radiation pressure prevents the necessary gravitational instability of baryons needed for structure growth. In MOG, according to Moffat \cite{moffat}, phions, which are the dominant particles of the universe in this epoch,  experience gravitational instability and plays the role of dark particles in GR. The evolution of phion perturbations can be compared with the dark matter perturbation in GR which is a logarithmic function of the scale factor in this epoch. 
During the stage between matter-radiation equality and photon decoupling, there exists phion dominated era. Photons and baryons still form a tightly coupled fluid. Therefore baryons again experience damped oscillations with smaller amplitude in comparison to those existed in radiation dominated epoch because of decreasing the effective sound velocity \cite{julienlesgourgues}. And phion perturbations have a growing behaviour which can be compared with a linear function of the scale factor which is the growing mode of dark matter perturbation after matter-radiation equality in GR. Well after decoupling, the baryons lose the pressure support of photons and gravitational instability starts in baryons and phions and we expect that baryon perturbations approach to the phion perturbations and both grow. The solution (\ref{w0}) is a background solution for this epoch.  After that, during matter domination, baryon perturbations grow and their evolution must be compared again with the scale factor.

Let's to consider a sequence of cosmological epochs with a single component fluid; radiation, phion or matter (For investigating the cosmological epochs in MOG using the phase space analysis see \cite{jamal}). Within any epoch we assume that the $G$-field density is also considerable. Starting from the early universe, and defining $f=\frac{\dot {G}}{G}$, for $G$-radiation dominated universe, the first and last terms in the parentheses of equation (\ref{eeqn1}) dominate, so equations (\ref{eqn3}), (\ref{eeqn1}) and (\ref{eeqn2}) yield respectively:
\begin{subequations}\label{appradiation}
\begin{align}
&3H^2-8\pi G\rho^{(r)}-3fH + \frac{f^2}{8}=0\\
&3\dot{H}+\frac{3fH}{2}+16\pi G\rho^{(r)}-\frac{9f^2}{8}=0\\
&f^2+\dot{f}+3fH=0
\end{align}
\end{subequations}
These equations lead to the following differential equation for $H$:
\begin{eqnarray}\label{radiationH}
&-24(2H+xH')\sqrt{H(2H+xH')}
\pm 6\sqrt{6}H(2H+xH')\pm\sqrt{6}x\left(xH'^2+H(5H'+xH'')\right)=0
\end{eqnarray}
where a prime denotes $d/dx$ and $x=a/a_{eq}$ in which $a_{eq}$ is the value of the scale factor at the time of matter-radiation equality. For upper sign, equation (\ref{radiationH}) has the following solutions:
\begin{align}\label{H1}
H&\sim  x^s\hspace{1 in}s= {9\pm2\sqrt{30}}, {-2}
\end{align}
 The value of $s= {9+2\sqrt{30}}$ is not acceptable because it leads to a contracting universe.
The lower sign in equation (\ref{radiationH}), gives also $s=-2$. Substituting (\ref{H1}) into (\ref{appradiation}c) and (\ref{appradiation}b), we find for $s=-2$:
\begin{align*}\label{H2}
H&\sim x^{-2}
\end{align*}
\begin{align}
G&=Const
\end{align}
\begin{align*}
\rho_r&\sim x^{-4}
\end{align*}
These are like the radiation-dominated case in standard cosmology. And for $s= {9-2\sqrt{30}}$:
\begin{align*}
H&\sim x^{9-2\sqrt{30}}=x^{-1.95}
\end{align*}
\begin{align}
G&\sim x^{-2\sqrt{6(11-2\sqrt{30}})}=x^{-1.045}
\end{align}
\begin{align*}
\rho_r&\sim x^{2(9+2\sqrt{30})+2\sqrt{6(11-2\sqrt{30})}}=x^{-2.86}
\end{align*}
For this solution we see that the strength of Newtonian gravitational constant decreases with the expansion of the universe in this epoch. Also the expansion rate of the universe increases and the radiation density decreases more quickly in comparison to the standard cosmology.  This behaviour of radiation density is not surprising, because here we deal with approximate cosmological equations in which only two dominant densities are involved and thus the conservation of energy-momentum tensor of ordinary matter does not necessarily hold. Moreover it is easy to check that for both solutions $\rho_{G}\sim \rho_r$, as we expected. In this era, since the neutral phion particles don't couple to photons, the right hand side of equation (\ref{eqn5}) is zero. Therefore the value of vector field is zero and from (\ref{eqn4}), $\mu=Const$, and thus $\rho_{\mu}\sim \rho_{\phi}\sim 0$.

A similar calculation can be performed for a universe dominated by $G$-phion or $G$-matter.  Here we shall only give the final results. For the first case, two solutions are:
\begin{align*}\label{in1}
H&\sim x^{-2}
\end{align*}
\begin{align}
G&\sim x^2
\end{align}
\begin{align*}
\rho_{\phi}&\sim x^{-6}
\end{align*}
and
\begin{align*}\label{in2}
H&\sim x^{9-2\sqrt{30}}=x^{-1.95}
\end{align*}
\begin{align}
G&\sim x^{2(1-\sqrt{5(11-2\sqrt{30})}=x^{1.04}}
\end{align}
\begin{align*}
\rho^{(\phi)}&\sim x^{4(4-\sqrt{30})+2\sqrt{5(11-2\sqrt{30})}}=x^{-4.95}
\end{align*}
In this epoch, according to these equations, the gravitational constant increases as the universe expands. Also $\rho_m$ and $\rho_{\mu}$ are ignorable compared to $\rho_{\phi}$. Thus from equation (\ref{eqn5}) we must have:
\begin{equation}
\rho_m\sim \left. \mu^2\phi_0\right|_{x\rightarrow 1}\ll\rho_{\phi}\sim \left.\mu^2 \phi_0^2\right|_{x\rightarrow 1}
\end{equation}
and 
\begin{equation}\label{c1}
\rho_\mu\sim \left. \frac{H^2}{G}(\frac{\mu'}{\mu})^2\right|_{x\rightarrow 1}\ll\rho_{\phi}\sim \left.\mu^2 \phi_0^2\right|_{x\rightarrow 1}
\end{equation}
These inequalities can be satisfied choosing the values of $\mu$ and $\phi_0$ large enough at equality time. We have not put any bound on the derivative of $\mu$. This becomes clear next when we discuss the $G$-matter era.

Finally for a $G$-matter dominated, one gets:
\begin{align*}\label{m1}
H(x)\sim & x^{-13/6}
\end{align*}
\begin{align}
G(x)\sim & x^{-4/3}
\end{align}
\begin{align*}
\rho_b\sim & x^{-3}.
\end{align*}
and
\begin{align*}\label{m2}
H&\sim x^{9-2\sqrt{30}}=x^{-1.95}
\end{align*}
\begin{align}
G&\sim x^{{\frac{2}{3}\left(-2+\sqrt{130+60(9-2\sqrt{30})}\right)}}=x^{{1.04}}
\end{align}
\begin{align*}
\rho^{(b)}&\sim x^{{\frac{4}{3}+2(9-2\sqrt{30})-\frac{2}{3}\sqrt{130+60(9-2\sqrt{30})}}}=x^{{-4.95}}
\end{align*}
From these solution we see that the gravitational constant can decrease or increase with expansion in this era. The expansion rate of the universe decreases and the matter density decreases more quickly in comparison to the standard cosmology. Also the solution (\ref{m2}) is the same as (\ref{in2}) for $G$-matter domination. It is natural since the phion and matter particles are both pressureless.
To determine the density of $\phi$ and $\mu$ fields, let us assume that the first one is ignorable in comparison to the second in equation (\ref{eqn4}). This leads:
\begin{equation}
(\frac{\dot{\mu}}{\mu})'+\frac{1}{x}(3-\frac{xG'}{G})\frac{\dot{\mu}}{\mu}=0
\end{equation}
Using (\ref{m1}) and (\ref{m2}), this equation can now be integrated analytically and has the solution:
\begin{equation}\label{m}
\frac{\dot{\mu}}{\mu}\sim 
\left \{
\begin{split}
x^{-13/3} & \hspace{1cm} s={-13/6 }\\
x^{{-1.96}} & \hspace{1cm} s=9-2\sqrt{3}
\end{split}
\right.
\end{equation}
This yields:
\begin{equation}
\frac{\rho^{(\mu)}}{\rho^{(b)}}\sim 
\left \{
\begin{split}
x^{-13/3} & \hspace{1cm} s={-13/6} \\
x^{{-0.01}} & \hspace{1cm} s=9-2\sqrt{3}
\end{split}
\right.
\end{equation}
which shows that for $G$-matter dominated epoch: $\rho^{(\mu)}\ll \rho^{(b)}$. Also from (\ref{m}):
\begin{equation}
\mu\sim 
\left \{
\begin{split}
\exp\left({-\frac{6}{13}\left .\frac{{\mu}'}{\mu}\right |_{x=1}x^{-13/6}}\right ) & \hspace{1cm} s={-13/6} \\
\exp\left({-\frac{1}{{0.01}}\left .\frac{{\mu}'}{\mu}\right |_{x=1}x^{{-0.01}}}\right ) & \hspace{1cm} s=9-2\sqrt{3}
\end{split}
\right.
\end{equation}
and thus:
\begin{equation}
\frac{\rho^{(\phi)}}{\rho^{(\mu)}}\sim 
\left \{
\begin{split}
x^{4/3}\exp\left({{\frac{12}{13}}\left .\frac{{\mu}'}{\mu}\right |_{x=1}x^{-13/6}}\right ) & \hspace{1cm} s={-13/6} \\
x^{{-4.94}}\exp\left({{200}\left .\frac{{\mu}'}{\mu}\right |_{x=1}x^{{-0.01}}}\right ) & \hspace{1cm} s=9-2\sqrt{3}
\end{split}
\right.
\end{equation}

Since in this era $x\gg 1$, we see that the positive values of $\frac{{\mu}'}{\mu}$ with large enough amplitude at equality time are needed to have $\rho_{\phi}\ll \rho_{\mu}$. Moreover  
the ratio of $\frac{\mu'}{\mu^2}$ must be small enough to be neglected compared with $\frac{\sqrt{G}\phi_0}{H}$ according to (\ref{c1}) which obtained before. Both of these conditions can be satisfied only in the case of sufficiently large $\mu$ at equality time. In this way, the universe experiences two stages of $G$-phion and $G$-matter dominated in MOG. Therefore we deduce that in the early universe, the vector field is very massive \cite{moffatarxiv2014}.
\section{Instability in an Expanding Universe}

To study the Newtonian growth of adiabatic perturbations we use the hydrodynamical equations of MOG theory. Since in FRW universe the energy-momentum tensor of matter is conserved, the continuity and Euler equations don't change. But this is not true for Poisson equation. This is because of coupling between matter and the vector field \cite{roshanabbassi}. In MOG theory the hydrodynamical equations are as follows

\emph{Continuity equation}

\begin{align}\label{continuityequation}
\frac{\partial \rho}{\partial t}+\boldsymbol\nabla\cdot (\rho\textbf{V} )=0
\end{align}

{\emph{Euler equations}

\begin{align}\label{eulerequations}
\frac{\partial \textbf{V}}{\partial t}+(\textbf{V}\cdot{\boldsymbol\nabla})\textbf{V}+\frac{\boldsymbol\nabla P}{\rho}+\boldsymbol\nabla\Phi=0
\end{align}

\emph{Modified Poisson equation}

\begin{align}\label{modifiedpoisson}
\nabla ^2\Phi=4\pi G_N\rho+\alpha \mu_0^2 G_N\int\frac{e^{-\mu_0\left | \textbf x-\textbf x'\right | }}{\left | \textbf x -\textbf x'\right | }\rho(\textbf x') d^3\textbf x'
\end{align}
where $\mu_0$ is the background value of $\mu$-field and $\alpha$ is a constant introduced in section 2.  It must be noted that the free parameters of MOG, $\alpha$ and $\mu_0$ (which is constant in the weak field approximation of this theory), are not universal and take different values for different galaxies \cite{rah1,rah2}. Also note that for $\alpha=0$ or $\mu _0=0$, the standard Poisson equation is recovered. Perturbing the background values of matter density, $\rho_0$, the gravitational potential, $\Phi_0$, and 3-velocities, $\textbf{V}$, of cosmic fluid as:
\begin{align}
\rho(\textbf{x},t)=\rho_0 (t)+\delta\rho(\textbf{x},t)\label{prho}\\
\textbf{V}(\textbf{x},t)=\textbf{V}_0(t)+\delta\textbf{V}(\textbf{x},t)\label{pvelocity}\\
\Phi(\textbf{x},t)=\Phi_0(t)+\delta \Phi(\textbf{x},t)\label{ppotential}
\end{align}
in which $\textbf{V}_0(t)$ obeys the Hubble law: $\textbf{V}_0(t)=H(t) \textbf{x}$. The remarkable point is that in MOG theory such as GR, there is no need to use the Jeans swindle \cite{binneytremaine} to study the Newtonians instability in an expanding universe. This may be easily understood from (\ref{modifiedpoisson}). Since the background density is only time dependent, integrating the last term of this equation shows that the gravitational acceleration is proportional to $\textbf{x}$ which is consistent with Hubble law.
Substituting (\ref{prho})-(\ref{ppotential}) into (\ref{continuityequation})-(\ref{modifiedpoisson}) and keeping only the linear terms, we obtain
 
\begin{align}
\frac{\partial \delta \rho}{\partial t}+\rho_0\boldsymbol\nabla\cdot(\delta\textbf{V})+{\boldsymbol\nabla}\cdot(\delta\rho\textbf{V}_0)=0
\label{con}
\end{align}
\begin{align}
\frac{\partial(\delta\textbf{V})}{\partial t}+(\textbf{V}_0\cdot{\boldsymbol\nabla})\delta\text{V}+(\delta\textbf{V}\cdot{\boldsymbol\nabla})\textbf{V}_0+\frac{c_s^2}{\rho_0}\boldsymbol\nabla\delta\rho+\nabla(\delta \Phi)=0
\end{align}
\begin{align}
\nabla ^2\delta\Phi=4\pi G_N\delta\rho+\alpha \mu_0^2 G_N\int\frac{e^{-\mu_0\left | \textbf x-\textbf x'\right | }}{\left | \textbf x -\textbf x'\right | }\delta \rho(\textbf x') d^3\textbf x'
\label{poi}
\end{align}
in which $c_s^2=\partial p/\partial \rho$ is the square of the speed of sound for adiabatic perturbation. For an expanding universe, it is convenient to use the  comoving coordinate defined as: 
\begin{align}
\textbf q=\frac{\textbf x}{a(t)}\\
\end{align}
the partial derivatives are related by: 
\begin{align}
\left(\frac{\partial}{\partial t}\right)_\textbf x = \left(\frac{\partial}{\partial t}\right)_\textbf q - \left( \textbf V_0 \cdot\boldsymbol\nabla_\textbf x \right)\\
\boldsymbol\nabla_\textbf x =\frac{1}{a}\boldsymbol\nabla_\textbf q
\end{align}
Introducing the fractional energy density perturbation, $\delta=\frac{\delta\rho}{\rho_0}$ in equations (\ref{con})-(\ref{poi}), we finally obtain:
\begin{align}\label{pcon}
\frac{\partial \delta }{\partial t}+\frac{1}{a}\boldsymbol\nabla.(\delta\textbf{V})=0
\end{align}
\begin{align}\label{peuler}
\frac{\partial\delta\textbf{V}}{\partial t}+H \delta\textbf{V} +\frac{c_s^2}{a}\boldsymbol\nabla\delta+\frac{1}{a}\boldsymbol\nabla(\delta \Phi)=0
\end{align}
\begin{align}\label{ppoisson}
\boldsymbol\nabla ^2\delta\Phi=4\pi  G_N\rho_0 a^2+\alpha \mu_0^2 G_N\rho_0a^4\int\frac{e^{-\mu_0a\left | \textbf q -\textbf q'\right | }}{\left | \textbf q -\textbf q'\right | }\delta (\textbf q') d^3\textbf q'
\end{align}
Taking the divergence of (\ref{peuler}) and using (\ref{pcon}) and (\ref{ppoisson}), we derive the following equation for the fractional amplitude of density perturbations:
\begin{align}\label{per1}
\ddot{\delta}+2H\dot{\delta}-\frac{c_s^2}{a^2} \nabla^2 \delta -4\pi G_N \rho_0\delta - \alpha \mu_0^2 G_N\rho_0a^2\int\frac{e^{-\mu_0 a \left | \textbf q -\textbf q'\right | }}{\left | \textbf q -\textbf q'\right | }\delta (\textbf q') d^3\textbf q'=0
\end{align}
Taking the Fourier transform $\delta (\textbf q, t)=\frac{1}{2\pi^{3/2}}\int \delta _k(t) e^{i\textbf k\cdot \textbf q}d^3\textbf k$, equation (\ref{per1}) therefore becomes
\begin{align}\label{per2}
\ddot{\delta_k}+2H\dot{\delta_k}+(\frac{c_s^2 k^2}{a^2} - 4\pi G_N\rho_0-\alpha \mu_0^2G_N\rho_0a^2 \int\frac{e^{-\mu_0 a \left | \textbf q -\textbf q'\right | +i\textbf k\cdot \left( \textbf q -\textbf q'\right)}}{\left | \textbf q -\textbf q'\right | } d^3\textbf q')\delta_k=0
\end{align}
Performing the above integral over all values of $\textbf q$ leads to:
\begin{align} \label{per3}
\ddot{\delta_k}+2H\dot{\delta_k}+\left[ \frac{c_s^2}{a^2} k^2 -4\pi G_N \rho_0\left ( 1+ \frac{\alpha\mu_0^2 a^2}{k^2+ \mu_0^2a^2}\right)\right]\delta_k = 0 
\end{align}
Thus the generalized physical Jeans length in MOG is:
\begin{eqnarray}\label{jeans}
\lambda_J^{phys.}=\frac{2\pi a}{\tilde{k_J}}
\end{eqnarray}
in which $\tilde{k_J}^2=\frac{{k_J}^2}{2}\left[(1-\frac{\mu_0 ^2}{k_J^2})+\sqrt{(1+\frac{\mu_0 ^2}{k_J^2})^2+\frac{4\alpha\mu_0 ^2}{k_J^2}}\right]$ and $k_J^2=\frac{4\pi G_N\rho_0}{c_s^2}$ are the corresponding wave number in MOG and GR respectively. It is clear that $\tilde{k_J}>{k_J}$, thus the perturbations on the scale  $\tilde{\lambda_J}<\lambda<\lambda_J$ describe the growing modes in MOG while those are sound waves in GR \cite{roshanabbassi}.
We see that for long wavelength perturbations, when $k^2\ll \tilde{k}_J^2$
equation (\ref{per3}) takes the following form:
\begin{eqnarray}\label{per4}
\ddot{\delta_k}+2H\dot{\delta_k}-4\pi G_N\rho_0(1+\alpha)\delta_k=0
\end{eqnarray}
which is modified by a factor of $(1+\alpha)$ as compared to the corresponding equation in general relativity. In terms of derivatives with
respect to the scale factor, equation (\ref{per4}) can be written as:
\begin{eqnarray}\label{cosmologicalperturbation}
\delta_k^{''}+(\frac{H'}{H}+\frac{3}{a}) \delta_k '-\frac{4\pi G_N\rho_0(1+\alpha)}{(aH)^2}\delta_k=0
\label{pera}
\end{eqnarray} 
On the other hand, in the small scale limit, the equation (\ref{per3}) reduces to the corresponding one in GR. But the evolution of acoustic perturbation is different in MOG theory since the corresponding Hubble parameter changes. 

In the next section we discuss the cosmological perturbation on scales much larger than the modified Jeans scale for the background space-times derived in section 3.

\section{Cosmological perturbation}
In the following, considering the background solutions of section 3,  we obtain the analytical solutions of matter fluctuations above the modified Jeans length. We concern with $G$-radiation, $G$-phion and  $G$-matter dominated epochs.  
 \begin{itemize}

\item{$G$-radiation dominated era}

As noted earlier, before the last scattering, baryons are coupled to photons whose pressure supports oscillation on this scale. Thus there is no gravitational instability in baryon component and only phion perturbation can grow. Moreover in this epoch, the phion density is ignorable in comparison to the radiation density, thus the last term in equation (\ref{cosmologicalperturbation}) can be neglected. substituting the bachground solution (\ref{H2}) in (\ref{cosmologicalperturbation}) leads to:
\begin{eqnarray}\label{radiationdominated1}
\delta''_{k\phi}+\frac{1}{x}\delta'_{k\phi}=0
\end{eqnarray}
Therefore, the cosmological perturbations of phion particles in radiation dominated epoch evolves as:
\begin{eqnarray}\label{radiationdominated2}
\delta_{k\phi}(x)=C_1 +C_2\ln x
\end{eqnarray}
which is just what we would have expected \cite{mukhanov} since phion particles in MOG play the role of dark matter particles in general relativity. The other solution (\ref{H1}) doesn't lead to any growing density perturbation mode. 
\item{$G$-phion dominated era}

According to Moffat \cite{moffatarxiv2015},  between epochs of equality and last scattering, the universe is dominated by phion particles and their fluctuations have a dominant contribution in structure formation. To find the growing mode of phion perturbation, we substitute the expressions (\ref{in1}) and (\ref{in2}) into equation (\ref{cosmologicalperturbation}). This gives the following general solutions respectively:
\begin{align}
\delta_{k\phi}=C_1K_0(\frac{\sqrt{A}}{x})+C_2I_0(\frac{\sqrt{A}}{x})
\end{align}
\begin{align}
\delta_{k\phi}=(\frac{1}{x})^{1/40}\left(C_1I_{0.048}(1.92\sqrt{A}x^{13/25})+C_2I_{-0.048}(1.92\sqrt{A}x^{13/25}\right).
\end{align}
where $A=\frac{4\pi G_N(1+\alpha)\rho_{eq}^{(\phi)}}{H_{eq}^2}$ and $I_j$ and $K_i$  are  the modified Bessel functions of the first and second kind of order $i$, respectively. Knowing the fact that this epoch occurs near $a\sim a_{eq}$, we can write $x=1+\bar{x}$  and so the above expressions can be expanded in terms of $\bar{x}$ by the Taylor expansion of Bessel function \cite{arf}. In the limit of small $\bar{x}$, both of these functions reduce to a linear function of $\bar{x}$. This is just what we would have expected for this epoch.

\item{Matter-phion perturbations}

After decoupling, it is expected that the baryonic fluctuations grow and finally catch up with the phion perturbation. To investigate the evolution of these perturbations, let us generalize equation (\ref{cosmologicalperturbation}) to the case that the cosmic fluid contains several components interacting gravitationally:
\begin{eqnarray}\label{multicomponents}
\delta_{ki}^{''}+(\frac{H'}{H}+\frac{3}{x}) \delta_{ki} '-\frac{4\pi G_N(1+\alpha )\rho_0^{(j)}}{(xH)^2}\delta_{kj}=0
\end{eqnarray}
where $j$ runs over all components. Thus we obtain:
\begin{eqnarray}\label{phionbaryonperturbations1}
\begin{split}
&\delta_{kb}^{''}+(\frac{H'}{H}+\frac{3}{x}) \delta_{kb} '-\frac{4\pi G_N(1+\alpha )}{(xH)^2}\big(\rho_0^{(b)}\delta_{kb}+\rho_0^{(\phi)}\delta_{k\phi}\big)=0\\&
\delta_{k\phi}^{''}+(\frac{H'}{H}+\frac{3}{x}) \delta_{k\phi} '-\frac{4\pi G_N(1+\alpha )}{(xH)^2}\big(\rho_0^{(b)}\delta_{kb}+\rho_0^{(\phi)}\delta_{k\phi}\big)=0
\end{split}
\end{eqnarray}
Taking 
\begin{eqnarray}\label{delta}
\begin{split}
&\Delta_k=\delta_{kb}-\delta_{k\phi}\\&
\rho_0^{(m)}\delta _{km}=\rho_0^{(b)}\delta_{kb}+\rho_0^{(\phi)}\delta_{k\phi}
\end{split}
\end{eqnarray}
where $\rho_0^{(m)}=\rho_0^{(b)}+\rho_0^{(\phi)}$ is the total density of pressureless matter, we can decouple equations (\ref{phionbaryonperturbations1}) to obtain
\begin{eqnarray}\label{dec}
\begin{split}
&\Delta_k^{''}+(\frac{H'}{H}+\frac{3}{x})\Delta_k '=0\\&
\delta_{km}^{''}+(\frac{H'}{H}+\frac{3}{x}) \delta_{km} '-\frac{4\pi G_N(1+\alpha )}{(xH)^2}\rho_0^{(m)}\delta_{km}=0
\end{split}
\end{eqnarray}
Substituting the background solution which we have found earlier, equation (\ref{w0}), we finally obtain:
\begin{eqnarray}\label{dec}
\begin{split}
&\Delta_{\phi b}=D_1+D_2x^{-\frac{1}{12}}\\&
\delta_{km}=x^{-1/24}\left(C_1 I_{-0.1}\left(13.23 x^{5/12}\right)+ C_2 I_{0.1}\left(13.23 x^{5/12}\right)\right)
\end{split}
\end{eqnarray}
in which we have used the present values of quantities as initial conditions.
Now Consider 
\begin{eqnarray}\label{phionbaryonperturbations4}
\frac{\delta_{kb}}{\delta_{k\phi}}=\frac{\rho_0^{(m)}\delta_{km}+\rho_0^{(\phi)}\Delta_{\phi b}}{\rho_0^{(m)}\delta_{km}-\rho_0^{(b)}\Delta_{\phi b}}
\end{eqnarray}
and also note that according to (\ref{dec}), both modes of $\delta_{km}$ are growing in contrast to $\Delta_{\phi b}$. Thus, we conclude that $\delta_{kb}$ approches $\delta_{k\phi}$ during this epoch.

\item{$G$-matter dominated era}

In this era, such as the previous one, it is more convenient to use the present values of quantities as initial conditions. Now by substituting both solutions (\ref{m1}) and (\ref{m2}) into (\ref{cosmologicalperturbation}), we obtain the following perturbation equations:
\begin{align}
\delta_{k}''+\frac{5}{6x}\delta_k'-\frac{3}{2}\bar{\Omega} _b^0(1+\alpha )x^{-2/3}\delta =0 
\end{align} 
\begin{align*}
\delta_k''+\frac{1.04}{x}\delta_k'-\frac{3}{2}\bar{\Omega} _b^0(1+\alpha )x^{-3.04}\delta =0 
\end{align*}
which correspond to the two background solutions (\ref{m1}) and (\ref{m2}) respectively and $\bar{\Omega} _b^0=\left .\frac{8\pi G_N \rho_0^{(b)}}{H^2}\right |_{x=1}$ is the present density parameter of baryons. It follows that:
\begin{align}
\delta_k(x)=x^{1/12} \left(C_1 I_{-0.125}( 1.15 x^{2/3}) + C_2 I_{0.125} (1.15 x^{2/3})\right).
\end{align} 
\begin{align*}
\delta_k(x)=x^{-1/50}\left(C_{1} I _{0.03} (1.47 x^{13/25})+
C_{2}I _{-0.03} (1.47 x^{13/25})\right)
\end{align*}
The large argument asymptotic expansions of modified Bessel functions may be written as \cite{arf},\cite{jadid}:
\begin{equation}
\label{sum}
I_{\nu}(z)\sim \frac{e^z}{(2\pi z)^{1/2}}\sum_{n=0}^{\infty}(-1)^n\frac{a_n(\nu)}{z^n}\pm i e^{\pm i\pi\nu} \frac{e^{-z}}{(2\pi z)^{1/2}}\sum_{n=0}^{\infty}\frac{a_n(\nu)}{z^n}
\end{equation}
where:
\begin{equation}
a_n(\nu)=(-1)^n\frac{\cos{\pi \nu}}{\pi }\frac{\Gamma (n+1/2+\nu)\Gamma (n+1/2-\nu)}{2^{n}\Gamma (n+1)}
\end{equation}
Thus for deep inside matter domination where the first term in summations of (\ref{sum}) is dominant and the perturbations grow as:
\begin{align}
\delta_k(x)=x^{-1/4}e^{1.15 x^{2/3}}
\end{align}
or
\begin{align*}
\delta_k(x)=x^{-14/50}e^{1.47 x^{13/25}}
\end{align*}
in MOG. Both of the above matter perturbations grow more quickly than $x$ which is the corresponding mode of perturbation for a universe occupied by matter in GR.

\end{itemize}
\section{Conclusion}
In this paper we have applied the Newtonian treatment of cosmological perturbations to MOG. To get some self-consistent background solutions, we consider two components for the universe and assume that the universe is occupied by a dynamical gravitational constant together with a cosmic fluid density. These solutions are explored analytically with respect to the scale factor. We have found the following evolutionary stages for the cosmological perturbations. First, a logarithmic growth in $G$-radiation epoch, followed by a linear growth in $G$-phion epoch, then followed by approaching the matter perturbation to the phion perturbation and finally an exponential growth in $G$-matter epoch.

We have seen that the gravitational instability in the Newtonian limit of MOG has two aspects. First, there are some perturbations on the small scales which lead to stable modes in GR but unstable modes in MOG and thus contribute in the structure formation. The other more important aspect is providing a mechanism  for producing faster growing perturbations than GR especially during recombination era, the so-called phion dominated era in MOG. This is a necessary condition for any gravitational theory explaining the cosmic structure without introducing the dark matter and found by Skordis, et al \cite{sko}
in TeVeS theory \cite{tev}. Moreover in any alternative of GR, the power spectrum of fluctuations must be compared with the observational power spectrum of galaxies \cite{dodel}.
 Other common interesting feature of TeVeS and MOG is that in both theories, the vector field plays a significant role in enhancing the growth of structures \cite{sko, dodel,mar}. 

Here these results are obtained in the limit of sub-horizon scales and would be more clear by considering the relativistic viewpoint of gravitational instability in MOG theory. 
\[ \]
{\bf Acknowledgements:}

The authors would like to thank M. Roshan for the precise reading of the paper and providing us with his helpful comments. This work is  supported by a grant from university of Tehran.

\end{document}